\documentclass[10pt, conference, letterpaper]{IEEEtran}

\usepackage{algorithm}    % include the algorithm and table package
\usepackage{algpseudocode}
\usepackage{amsmath,bm}
\usepackage{amssymb}    % ÓÃÀ´ÅÅ°æƯÁÁµÄÊýѧ¹«Ê½
\usepackage{mathrsfs}   % Ó¢ÎÄ»¨Ìå×ÖÌå
\usepackage{graphicx,subfigure}

\usepackage{fancyhdr}  % ÉèÖÃҳüҳ½Å
\usepackage{cases}

\makeatletter
\renewcommand\normalsize{%
   \@setfontsize\normalsize\@xpt\@xiipt
   \abovedisplayskip 1\p@ \@plus1\p@ \@minus2\p@
   \abovedisplayshortskip \z@ \@plus1\p@
   \belowdisplayshortskip 5\p@ \@plus1\p@ \@minus2\p@
   \belowdisplayskip \abovedisplayskip
   \let\@listi\@listI}
\makeatother

\ifCLASSINFOpdf
\else
\fi
\begin{document}
\title{A LSE and Sparse Message Passing-Based Channel Estimation for mmWave MIMO Systems }
%\title{A Novel Channel Estimation Approach for Millimeter Wave MIMO Systems based on LSE and Sparse Message Passing }
%
%
%% author names and affiliations
%% use a multiple column layout for up to three different
%% affiliations
%\author{\IEEEauthorblockN{Chongwen \underline{ }Huang}
%\IEEEauthorblockA{Singapore University of \\Technology and Design\\
%Singapore, 487372 \\
%Email: huang@mymail.sutd.edu.sg}
%\and
%\IEEEauthorblockN{Lei Liu}
%\IEEEauthorblockA{Twentieth Century Fox\\
%Singapore, 487372 \\
%Email: lliu_0@stu.xidian.edu.cn}
%\and
%\IEEEauthorblockN{James Kirk\\ and Montgomery Scott}
%\IEEEauthorblockA{Starfleet Academy\\
%San Francisco, California 96678--2391\\

\author{\IEEEauthorblockN{Chongwen Huang\IEEEauthorrefmark{1}, Lei Liu\IEEEauthorrefmark{2}, Chau Yuen\IEEEauthorrefmark{1}, Sumei Sun\IEEEauthorrefmark{3} \\
\IEEEauthorrefmark{1}Singapore University of Technology and Design, 487372, Singapore\\
\IEEEauthorrefmark{2}State Key Lab of ISN, Xidian University, Xi'an, 710071, China\\
\IEEEauthorrefmark{3}Institute for Infocomm Research (I2R), A$^{\star}$STAR, 138632, Singapore\\
E-mail: chongwen$\_$huang@mymail.sutd.edu.sg }
}
\maketitle
\begin{abstract}
In this paper, we propose a novel channel estimation algorithm based on the Least Square Estimation (LSE) and Sparse Message Passing algorithm (SMP), which is of special interest for Millimeter Wave (mmWave) systems, since this algorithm can leverage the inherent sparseness of the mmWave channel. Our proposed algorithm will iteratively detect exact the location and the value of non-zero entries of sparse channel vector without its prior knowledge of distribution. The SMP is used to detect exact the location of non-zero entries of the channel vector, while the LSE is used for estimating its value at each iteration. Then, the analysis of the Cramer-Rao Lower Bound (CRLB) of our proposed algorithm is given. Numerical experiments show that our proposed algorithm has much better performance than the existing sparse estimators (e.g. LASSO), especially when mmWave systems have massive antennas at both the transmitters and receivers. In addition, we also find that our proposed algorithm converges to the CRLB of the genie-aided estimation of sparse channels in just a few turbo iterations.
\end{abstract}
%The key to accurate estimation is the exploitation of the inherent sparsity of the channel. Our proposed estimator is based on Least Square Estimation (LSE) and Bernoulli Message Passing (SMP) algorithm, which iteratively detects the location and value of non-zero entities in the sparse channel matrix by leveraging the inherent sparseness of mmWave channel. Numerical experiments show that it has far better performance (approach four order of magnitudes) than the conventional LSE estimator, especially when mm wave system has massive antennas at both the transmit and receive sides. In addition, we also find that this algorithm only needs three turbo iterations to achieve convergence.

% Firstly, we present an intermediate
%virtual channel representation that captures the essence of physical modeling and provides simple geometric interpretation of the scatting environment. Then, we consider the case that virtual channel matrix follows i.i.d. Bernoulli-Gaussian distribution
%
%Our proposed algorithm involves to iteratively detect exact the location and the value of non-zero entries in virtual channel matrix.
% \fi
\IEEEpeerreviewmaketitle
\vspace{-2mm}
\section{Introduction}
\vspace{-1mm}
Millimeter wave (mmWave) is receiving tremendous interest by academia, industry, and government for future 5G cellular systems \cite{rappaport_mmWave_model2013,SP_for_2016}. The main reason is that the majority of our current wireless communications systems operating in the microwave spectrum (i.e.,$ <$6 GHz) are by now crowded and limited bandwidth. MmWave can take full advantage of spectrum from 30 GHz to 300 GHz and provide beyond 2 GHz of bandwidth \cite{rappaport_mmWave_model2015}. However, the larger communication spectrum  means  more challenges. One of the main challenges, mmWave signal propagation is impaired by severe  signal attenuation. Recent urban model experiments show that path losses are 40 dB worse at 28 GHz compared to 2.8 GHz \cite{rappaport_mmWave_model2015}.

One way to overcome this severe signal attenuation of mmWave propagation paths is to improve beamforming gain by increasing number of transmit antennas and receive antennas. This leads to an important feature of mmWave channels, which is very sparse in both the angle and time domains \cite{SP_for_2016,P. Schniter_virtual}. Recently research measurements \cite{rappaport_mmWave_model2015} show that mmWave channels typically exhibit only 3-4 scattering clusters in dense-urban Non-Line-Of-Sight and Line-Of-Sight (LOS) environments. However, conventional MIMO channel estimation methods cannot be applied directly in mmWave systems since these methods did not account for mmWave channel sparsity. For example, in \cite{Y. Zhu_MP,wusheng_MP}, two joint channel estimation schemes for massive MIMO systems were proposed, and they were both based on message-passing iterative algorithms, which can reduce the complexity. However, they did not specialize for mmWave systems. This prompts the need to design efficient channel estimation techniques for the mmWave.

For the sparse channel estimation, a well-known approach is the LASSO \cite{LASSO} with a tuning parameter that trades between the sparsity and measurement-fidelity of the solution. In \cite{C. Carbonelli_sparse}, the authors investigated three algorithms for sparse channel estimation, and they were Approximate Maximum Likelihood Estimator (AMLD-SE), Iterative Detection/Estimation With Threshold (ITD-SE) and the Sphere Detection (SD-SE). In \cite{R. Niazadeh_sparse}, an alternating minimization method for sparse channel estimation was proposed. Simulation results showed that ITD-SE had a better performance than others. However, it needs an adaptive threshold selection strategy to keep the performance, which reduces its flexibility. Recently, approaches to deal with sparse channels have been proposed in literatures by resorting to Compressed Sensing (CS) theory. In \cite{Alk_channel_Estimation}, an adaptive CS algorithm for mmWave channel estimation was proposed. This algorithm is very suitable for the mmWave, while it relies on the design of a multi-resolution beamforming codebook.

In this paper, we propose a novel channel estimation scheme based on the Least Square Estimation (LSE) and Sparse Message Passing (SMP) algorithm. Compared with previously proposed sparse channel estimation, ours can yield a better performance since it not only can take full advantage of the inherent sparseness of mmWave channel, but also can leverage the virtues of LSE and SMP algorithm. Firstly, we introduce an intermediate virtual channel representation that is attractive since it can capture the essence of physical modeling and provides simple geometric interpretation of the scatting environment([see \cite{P. Schniter_virtual,Sayeed_SP2002}]). Then, our proposed algorithm involves to iteratively detect exact the location and the value of non-zero entries. The SMP is used to detect exact the location of non-zero entries of the channel vector, while the LSE is used for estimating its value at each iteration. Furthermore, we give the analysis of the Cramer-Rao Lower Bound (CRLB) of our proposed algorithm. Numerical simulations show that our algorithm exhibits far better performance than the conventional LSE estimator as well as existing sparse estimator (e.g. LASSO). In addition, we also find that this algorithm needs only four turbo iterations to nearly achieve the CRLB.

\textit{Notation}: $ a $ is a scalar, $ \mathbf{a} $ is a vector and $ \mathbf{A} $  is a matrix. $ \mathbf{A^T}$, $ \mathbf{A^H}$, $ \mathbf{A^{-1}}$, $ \mathbf{A^\dag}$ and $ \|\mathbf{A}\|_F $  represent transpose, Hermitian (conjugate transpose), inverse, pseudo-inverse and Frobenius norm of a matrix $ \mathbf{A} $, respectively. $ \mathbf{A} \otimes \mathbf{B} $ denotes the Kronecker product of $ \mathbf{A} $ and $ \mathbf{B} $, and $ vec(\mathbf{A}) $  is a vector stacking all the columns of $ \mathbf{A} $. $ diag(\mathbf{a}) $ is a diagonal matrix with the entries of $ \mathbf{a} $ on its diagonal, and $ diag(\mathbf{A}) $ is a vector that its entries come from the diagonal elements of $ \mathbf{A} $.
%
%In \cite{ A. Alkhateeb2014}, a channel estimation for the mmWave channel estimation problem based on adaptive CS was proposed, which enabled hybrid precoding to approach the performance of the digital precoding algorithms. However, it depends on the design of sparse hierarchical codebook.
\vspace{-1.5mm}
\section{SYSTEM MODEL}
\vspace{-1.5mm}
We consider a mmWave communication system that has $ N_t $ transmit antennas, $ N_r $  receive antennas, and the narrow band baseband received signal can be written as follows:
\begin{equation}\label{1}
  \mathbf{y}=\mathbf{Hs}+\mathbf{n}
\end{equation}
where $ \mathbf{H }\in \mathbb{C}^{N_r \times N_t}$ is the channel matrix, $ \mathbf{s} \in \mathbb{C}^{N_t \times 1}$ is the transmitted signal, $ \mathbf{y }\in \mathbb{C}^{N_r \times 1}$ is the received signal, and $ \mathbf{n}\sim \mathcal{N}(0,\sigma^2\mathbf{I}) $ is the Gaussian noise corrupting the received signal.
Since mmWave channels are expected to have limited scattering, we adopt a geometric channel model with $L$ scatterers. Each scatterer is further
assumed to contribute a single propagation path between transmitters and receivers \cite{rappaport_mmWave_model2015}. Under this model, the channel  $ \mathbf{H } $ can be expressed as:
\begin{equation}\label{2}
\mathbf{ H}=\sum_{l=1}^L\alpha_l\mathbf{a}_{r}(\theta_l)\mathbf{a}_{t}^H(\phi_l)
\end{equation}
where $ \alpha_l$ is the gain of the $ l $th path, $ \phi_l\in[0,2\pi]$, and $ \theta_l\in[0,2\pi]$ denote the $l$th path's azimuth angles of departure and arrival of transmitters and receivers respectively. Finally, $ \mathbf{a}_{r}(\theta_l)$ and $ \mathbf{a}_{t}(\phi_l)$ are the antenna array response vectors at receivers and transmitters respectively. If A Uniform Planar Array (UPA) is used, $ \mathbf{a}_{t}(\phi_l) $ can be written as:
\begin{equation}\label{3}
\!\!\!\!\mathbf{\mathbf{a}}_{t}(\phi_l)\!\!=\!\!\frac{1}{\sqrt{N_{t}}}[1,e^j\frac{2\pi}{\lambda}sin(\phi_l),...,e^j(N_{t}-1)\frac{2\pi}{\lambda}sin(\phi_l)]^T
\end{equation}
where $ \lambda$ is the signal wavelength, and $ d $ is the distance between antenna elements. The array response vectors at the receivers, $ \mathbf{a}_{r}(\theta_l)$, can be written in a similar fashion. Then, the channel can be written in a more compact form as:
\begin{equation}\label{4}
  \mathbf{H}=\mathbf{A}_{r}diag( \bm{\alpha})\mathbf{A}_{t}^H
\end{equation}
where $ \bm{\alpha} =\sqrt{\frac{N_{r}N_{t}} {\rho}}[\alpha_1,\alpha_2,...,\alpha_l]^T.$ The matrices \\
\begin{equation}\label{5}
  \mathbf{A}_{t}=[\mathbf{a}_{t}(\phi_1),\mathbf{a}_{t}(\phi_2),...,\mathbf{a}_{t}(\phi_l)]
\end{equation}
and
\begin{equation}\label{6}
\mathbf{A}_{r}=[\mathbf{a}_{r}(\theta_1),\mathbf{a}_{r}(\theta_2),...,\mathbf{a}_{r}(\theta_l)]
\end{equation} contain the transmitter and receiver array response vectors. Finally, $ \rho$ is the path-loss between the transmitters and receivers.

The mmWave channel is usually dominated by the LOS, or consists of a single reflected path \cite{J. Mo2014}. Large antenna arrays are deployed to get beamforming gain in order to combat the high path loss. Hence, we usually have $ L \ll min \{N_{r},N_{t}\}$.  %\cite{A.Ghosh_LOS}

Then, we adopt an intermediate virtual channel representation that keeps the essence of physical modeling without its complexity, provides a tractable linear channel characterization, and offers a simple and transparent interpretation on the effects of scattering and array characteristics on channel capacity. The finite dimensionality of the signal space allows the virtual channel model to be expressed as:
\begin{equation}\label{7}
\begin{split}
\mathbf{H}_v=\mathbf{W}_{r}^H\mathbf{H}\mathbf{W}_{t}
%&=\underbrace{[\alpha(\frac{-\tilde{N_r}}{N_r}),\alpha(\frac{-\tilde{N_r}+1}{N_r}),...,\alpha(\frac{\tilde{N_r}}{N_r})]^H}_{\mathbf{W}_{r}^H} \\
%&\times \mathbf{H} \underbrace{[\alpha(\frac{-\tilde{N_t}}{N_t}),\alpha(\frac{-\tilde{N_t}+1}{N_t}),...,\alpha(\frac{\tilde{N_t}}{N_t})] }_{\mathbf{W}_{t}}
\end{split}
\end{equation}
%Where $ \tilde{N_r}:=\frac{N_r-1}{2}$ and $ \tilde{N_t}:=\frac{N_t-1}{2}$ (assuming the $N_r$ and $N_t$ are odd), and where $\mathbf{H}_v \in \mathbb{C}^{N_r\times N_t}$ is no longer diagonal

Note that $ \mathbf{W}_{r} \in \mathbb{C}^{N_r \times N_r }$ and $ \mathbf{W}_{t} \in \mathbb{C}^{N_t \times N_t }$ are channel-invariant unitary Discrete Fourier Transform (DFT) matrices \cite{P. Schniter_virtual_sparse}. We recast (1) by the virtual channel representation as:
\begin{equation}\label{8}
\begin{split}
\mathbf{y}&=\mathbf{W}_{r}\underbrace{\mathbf{W}_{r}^H\mathbf{H}\mathbf{W}_{t}}_{:=\mathbf{H}_v}\mathbf{W}_{t}^{H}\mathbf{s}+\mathbf{n} \\
&=\mathbf{W}_{r}\mathbf{H}_v\mathbf{W}_{t}^{H}\mathbf{s}+\mathbf{n}
\end{split}
\end{equation}

\begin{figure}
  \begin{center}
  \includegraphics[width=80mm]{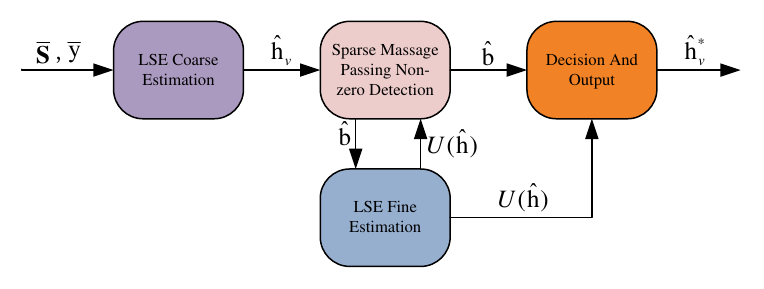} \vspace{-5mm} \\  %
  %\captionsetup{justification=centering}
  \caption{The processing for proposed LSE-SMP algorithm, which consists of three phases: Coarse Estimation, Sparse Message Passing Detection, and Fine Estimation.}
fig:env  \label{fig:env}\vspace{-8mm}
  \end{center}
\end{figure}\vspace{-0mm}
The fourier transformation $ \mathbf{W}_{t}$ and $ \mathbf{W}_{r}$  can be seen as a mapping from the antenna domain onto an angular domain and the entries of the matrix $ \mathbf{H}_v$ can be interpreted as the channel gains between $ N_t$ transmit and the $ N_r $ receive beams. In this paper, we assume that the array response vectors are along the directions defined in the DFT matrices. This means that $ \mathbf{H}_v$  is  perfectly sparse, i.e., has exactly $L$ non-zero entries.

Assuming the channel is time-invariant in the blocks ${t \in \{1,...,T \} }$. Then, $\mathbf{Y}\triangleq[\mathbf{y}_{1},...,\mathbf{y}_{T}] $, $\mathbf{S}\triangleq[\mathbf{\mathbf{s}}_1,...,\mathbf{s}_T] $
, and $\mathbf{N}\triangleq[\mathbf{\mathbf{n}}_{1},...,\mathbf{n}_{T}] $. We rewritten the channel model as:
\begin{equation}\label{9}
\begin{split}
\mathbf{Y} =\mathbf{W}_{r}\mathbf{H}_v\mathbf{W}_{t}^{H}\mathbf{S}+\mathbf{N}
\end{split}
\end{equation}
where $ \mathbf{Y}\in \mathbb{C}^{N_r \times T} $, $ \mathbf{S} \in \mathbb{C}^{N_t \times T} $, and $ \mathbf{N} \in \mathbb{C}^{N_r \times T} $. We define $ \mathbf{X}\triangleq \mathbf{W}_{t}^{H} \mathbf{S}$. Then, we recast the (9) as:
\begin{equation}\label{10}
\begin{split}
\mathbf{Y} =\mathbf{\mathbf{W}}_{r}\mathbf{H}_v\mathbf{X}+\bm{\mathbf{N}}
\end{split}
\end{equation}
Then, we can factorize the (10) \cite{Alk_channel_Estimation} as:
\begin{equation}\label{11}
\begin{split}
vec(\mathbf{Y})&=vec(\mathbf{W}_{r}\mathbf{H}_v\mathbf{X} + \mathbf{N})\\
& \stackrel{(a)}{=}(\mathbf{X}^T \otimes \mathbf{W}_{r} )vec(\mathbf{H}_v)+vec( \mathbf{N})
\end{split}
\end{equation}
where (a) follows from the equality $ vec(\mathbf{ABC})=(\mathbf{C^T} \otimes \mathbf{A})vec(\mathbf{B}) $. By defining the $ \mathbf{\bar{S}}\triangleq \mathbf{X}^T \otimes \mathbf{W}_{r} $, we simply the (11) in a compact fashion.
\begin{equation}\label{12}
\begin{split}
\mathbf{\bar{y}}=\mathbf{\bar{S}}\mathbf{h}_v+\mathbf{\bar{n}}
\end{split}
\end{equation}
where $ \mathbf{\bar{y}} \in \mathbb{C}^{N_rT \times 1} $, $ \mathbf{\bar{S}} \in \mathbb{C}^{N_rT \times N_tN_r} $, $ \mathbf{h}_v \in \mathbb{C}^{N_tN_r \times 1} $ and $ \mathbf{\bar{n}} \in \mathbb{C}^{N_rT \times 1} $.
The mmWave channel estimation problem is simplified to estimate virtual channel vector $ \mathbf{h}_v $ by giving the equivalent training matrix $ \mathbf{\bar{S}} $ and the observed signal $ \mathbf{\bar{y}} $.
%In the following section, we propose a novel algorithm to estimate exact positions and values of these non-zero entries in virtual channel vector $ \mathbf{h}_v $.
\begin{figure*}
  \begin{center}
  \includegraphics[width=155mm]{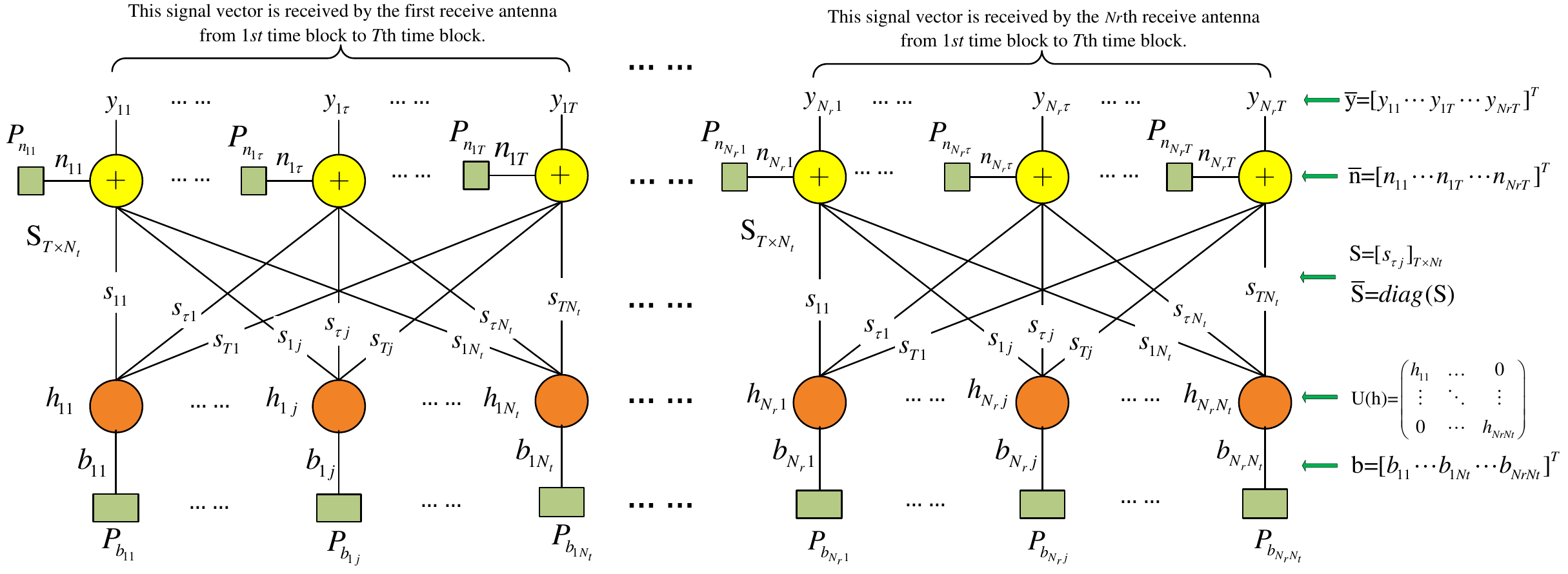} \vspace{-4mm}  \\%
  %\captionsetup{justification=centering}
  \caption{Factor Graph Representation for Proposed Sparse Message Passing (SMP) Detection Algorithm }

  \label{fig:env} \vspace{-8mm}
  \end{center}
\end{figure*}
\section{SPARSE CHANNEL ESTIMATION}
In this section, we  present a novel channel estimation algorithm based on the LSE and SMP algorithm, which is named LSE-SMP, and it consists of three phases: Coarse Estimation, Message Passing Detection, and Fine Estimation. This algorithm is shown in the Fig. 1. Since there is no priori knowledge of $ \mathbf{h}_v$, we initially adopt LSE method to obtain its coarse estimation. Then, we consider the estimation of non-zero positions in the channel vector $ \mathbf{h}_v $ as a detection problem, and propose a SMP algorithm to find these non-zero positions. Again, we apply LSE method by leveraging the estimated non-zero positions to obtain the fine estimation. The second step and third step will repeat until we obtain the steady estimation.
\subsection{LSE Coarse Estimation}
To find Coarse Estimation $ \mathbf{\hat{h}}_v $ based on the observed vector $\mathbf{\bar{y}}$, with a Mean Square Error (MSE), we can solve the following Least Square (LS) problem:
\begin{equation}\label{13}
\begin{split}
\mathbf{\hat{h}}_v &=arg \min_{\mathbf{h}_v}\|\mathbf{\bar{y}}-\mathbf{\bar{S}}\mathbf{h}_v\|^{2}_2\\
&=[\mathbf{\bar{S}}^T\mathbf{\bar{S}}]^{-1}\mathbf{\bar{S}}^T\mathbf{\bar{y}}
\end{split}
\end{equation}
It is noted that it is optimal estimator in the sense of MSE when the estimator does not have any prior knowledge about neither the sparsity structure of $ \mathbf{h}_v$ (i.e. the location of non-zero entries), nor its degree of sparsity (i.e. $L$ ).
\subsection{Sparse Message Passing Algorithm}
After we get the Coarse Estimation of $ \mathbf{h}_v $, we propose a fast iterative algorithm to find the positions of non-zero entries. This algorithm is named Sparse Message Passing since it can take full advantage of channel sparsity and message passing.
\subsubsection{Factor Graph Representation of mmWave Channel}
In order to get better understanding of our proposed algorithm, we show factor graph representation of channel vector $ \mathbf{h}_v $ in the following. Firstly, we decompose the $ \mathbf{h}_v $ into a diagonal coefficient matrix $ \mathbf{U}(h_{ij})  ( i \in [{0,...,N_r}],  j \in [{0,...,N_t}] ) $ and a column array $ \textbf{b} $. The column array $ \textbf{b}=[b_{ij} ]_{N_rN_t \times 1} $ is called position vector, and it represents the positions of non-zero in virtual channel vector $ \mathbf{h}_v $. The $ b_{ij}\in\{1,0\}$ can be seen as a Bernoulli distribution. Then, the $ \mathbf{h}_v $ can be recast as:
\begin{equation}\label{14}
\begin{split}
\mathbf{h}_v
 = \underbrace{ \left[\begin{array}{ccccc}
    h_{11}  &           &          &      &   0          \\
            &   \ddots  &          &                 \\
            &           & h_{1N_t} &                \\
            &           &          &    \ddots   &     \\
     0      &           &          &              &   h_{N_rN_t}      \\
\end{array}\right]}_{=\mathbf{U(h)}}
\underbrace{\left[\begin{array}{cccc}
  b_{11} \\
  \vdots \\
  b_{1N_t} \\
  \vdots \\
  b_{N_rN_t}
\end{array}\right]}_{=\mathbf{b}}
\end{split}
\end{equation}

The equivalent training matrix $ \mathbf{\bar{S}} $ can be expressed as the following block diagonal matrix by designing $\mathbf{X} $ and $ \mathbf{W}_r$.
\begin{equation}\label{15}
\begin{split}
\mathbf{\bar{S}}=diag(\mathbf{S})
%=  \left[\begin{array}{ccc}
%         \bm{\overline{S}_i}   &           &   0          \\
%            &   \ddots  &               \\
%         0  &           &  \bm{\overline{S}_i}            \\
%\end{array}\right]
\end{split}
\end{equation}
where $ \mathbf{\bar{S}} \in \mathbb{C }^{N_rT \times NrN_t}$, and $ \mathbf{S} $ can be seen as a matrix of transmitted training sequences, which will be received by $ i$th  $ (i \in {[1,...,N_r]})$ receive antenna from 1 to $ T $th time block. It can be written as:
\begin{equation}\label{16}
\begin{split}
\mathbf{S}=[s_{\tau j}]_{T \times N_t}
\end{split}
\end{equation}
where $ s_{\tau j} $ represents the transmitted training sequences by $ j $th  $(j \in {[1,...,N_t]})$ transmit antenna in the $\tau$th  $ (\tau \in {[1,...,T]})$ time block.
Then, we rewrite (12) as:
\begin{equation}\label{17}
\begin{split}
\!\!\!\!\!\!\underbrace{\left[\begin{array}{ccccc}
  y_{11} &  \ldots  & y_{1T}  &  \ldots  &  y_{N_rT}
\end{array}\right]^T}_\mathbf{\bar{y}} &= \mathbf{\bar{S}}\mathbf{h}_v +\mathbf{\bar{n}} \\
                   &=\mathbf{\bar{S}}\mathbf{U(h)b}+\mathbf{\bar{n}}
\end{split}
\end{equation}
% where $ y_{ij} $ represents received data stream by $ ith $  $(i \in {[1,...,N_r]}) $ receive antenna in the $ jth $  $ (i \in {[1,...,T]}) $ time block.

According to the factor graph analysis rule, we plot a factor graph to represent above equations, and it is shown in the Fig. 2. The nodes ($ n_{11},...,n_{N_rT}$) and ($ h_{11},...,h_{N_rN_t}$) are named the sum and variable nodes respectively.

Our proposed SMP algorithm is considered for estimating positions of non-zero. It is similar to the belief propagation decoding process of low density parity check code, in which the output message called extrinsic information on each edge is calculated by the messages on the other edges that are connected with the same node \cite{Lei2016_TVT}.
\subsubsection{Message Update at Sum Nodes of SMP}
To analysis a sum node that is shown in the Fig. 3, we can know the received signal at the $ \tau $th time block by the $ i$th receive antenna, and it can be expressed as:
\begin{equation}\label{18} \vspace{-1mm}
\begin{split}
y_{i\tau}= \sum_{j=1}^{N_t}s_{\tau j}h_{ij}b_{ij}+n_{i\tau}
\end{split}
\end{equation} \vspace{-0mm}

From assumption of channel model in the above section, we know that there are  $ L $ non-zero in the vector of $ \mathbf{b} $. Assuming that $b_{ij}$ is independent and identically distributed (i.i.d.). Then, we can know the initial probability of the Bernoulli distribution, which can be denoted as:
\begin{equation}\label{19}
\left\{
\begin{array}{l}
p_0(b_{ij}=1)=\frac{L}{N_rN_t}, \\
p_0(b_{ij}=0)=1-\frac{L}{N_rN_t}.
\end{array}
\right.\quad
\end{equation}

According to the law of large numbers, when $ N_t $ goes very large, we can know the term $ \sum_{j=1}^{N_t}s_{\tau j}h_{ij}b_{ij} $ can be approximated as Gaussian distribution \cite{Lei2015}. When we compute the probability of $ p(b_{ij}=1) $ from  $ i\tau $ sum node to $  ij $ variable node, we consider the messages from other variable nodes to sum node $ i\tau $ as noise. This can be expressed as:
\begin{equation}\label{20}
\begin{split}
\!\!\!\!y_{i\tau}\!\!= \!\!\underbrace{s_{\tau j}h_{ij}p(b_{ij}=1)}_{desired} +\underbrace{\sum_{m\neq j}^{N_t}s_{\tau m}h_{im}p(b_{im}=1)+n_{i\tau}}_{{interference}}
\end{split}
\end{equation}

Furthermore, we can compute its mean value and variance. The message update at the sum nodes is given by:
\begin{equation}\label{21}   %%
\left\{
\begin{array}{l}
e^{s}_{i\tau\rightarrow ij}(k)=\sum\limits_{m\neq j}s_{\tau m}\hat{h}_{im}p^{v}_{im \rightarrow i\tau}(k-1),\\
v^{s}_{i\tau\rightarrow ij}(k)={\sum\limits_{m \neq j}}s^{2}_{\tau m}p^{v}_{im \rightarrow i\tau}(k-1) \\ \times (v_{h_{im}}+\hat{h}_{im}^{2}(1-p^{v}_{im \rightarrow i\tau}(k-1)))+\sigma^{2}_{n}.
\end{array}
\right.
\end{equation}
where $ j,m\in \{1,2,...,N_t\}$, $\sigma^{2}_{n}$ is the variance of the Gaussian noise. In addtion, $ p^{v}_{im \rightarrow i\tau}(k-1) $ denotes the probability message of passing from the $ im $ variable node to $  i\tau $ sum node at $ (k-1)$th iteration \cite{Globalcom2016}. Similarly, $ e^{s}_{i\tau\rightarrow ij}(k) $ and $ v^{s}_{i\tau\rightarrow ij}(k)$ denote the mean and variance of $ s_{\tau j}h_{ij}$ passing from the $ i\tau $ sum node to $ ij$ variable node at $ k $th iteration respectively. $ \hat{h}_{im} $ and $ v_{h_{im}} $ denote the mean and variance of  $ h_{im} $ that are estimated in the fine phase.

Once we get the mean and variance from the $ i\tau $ sum node to $  ij $ variable node, we can compute its statistical probability according to the $ s_{\tau j}h_{ij}p(b_{ij}=1)\sim \mathcal{N }(e^{s}_{i\tau\rightarrow ij}(k),v^{s}_{i\tau\rightarrow ij}(k))$. Firstly, we define the Gaussian probability density function as:
\begin{equation}\label{22}   %%
f(x|\mu,\sigma)=\frac{1}{\sqrt{2\pi}\sigma}e^{-\frac{(x-\mu)^2}{2\sigma^2}}
\end{equation}

Then, we can give the probability from the $ i\tau $ sum node to $  ij $ variable node as follow:
\begin{equation}\label{23}   %%
\!\!\!p^{s}_{i\tau\rightarrow ij}(k) \!\!=\!\!\frac{1}{1+\frac{f(y_{i\tau}|e^{s}_{i\tau\rightarrow ij}(k),v^{s}_{i\tau\rightarrow ij}(k))}{f(y_{i\tau}|e^{s}_{i\tau\rightarrow ij}(k)+s_{\tau j}\hat{h}_{ij},v^{s}_{i\tau\rightarrow ij}(k)+s^{2}_{\tau j}v_{h_{ij}})}}
\end{equation}

\subsubsection{Message Update at Variable Nodes of SMP}

In terms of the message update at variable nodes, we consider variable nodes as a broadcast process \cite{Lei2016_TVT} and the message update at these nodes is given by:
\begin{equation}\label{24}    %%
p^{v}_{ij \rightarrow i\tau}(k)= \frac{1}{1+\frac{\prod_{t \neq \tau}(1-p^{s}_{it \rightarrow ij}(k-1))\cdot p_0(b_{ij}=0)}{\prod_{t \neq \tau}p^{s}_{it \rightarrow ij}(k-1)\cdot p_0(b_{ij}=1)}}
\end{equation}
where $ t,\tau \in \{1,2,...,T\}$, and, $ p^{s}_{it \rightarrow ij}(k-1) $ denotes the probability message passing from the $ it $ sum node to $  ij $ variable node at $ (k-1)$th iterations.
\begin{figure}
  \begin{center}
  \includegraphics[width=85mm]{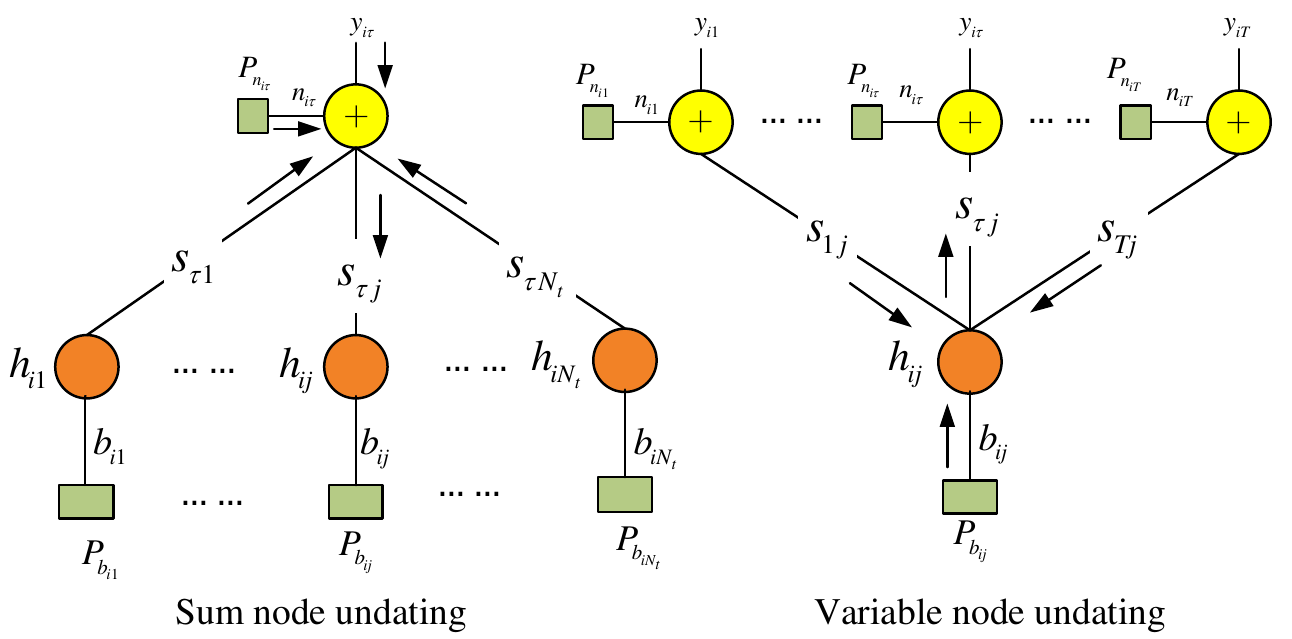} \vspace{-4mm}\\  %
  %\captionsetup{justification=centering}
  \caption{Message update at sum nodes and variable nodes. The output message called extrinsic information on each edge is calculated by the messages on the other edges that are connected with the same node. For the Gaussian-Bernoulli sparse signals, the messages passing on each edge are the mean and variance of a Gaussian distribution, and the non-zero probability of a Bernoulli distribution. }
  \label{fig:env} \vspace{-8mm}
  \end{center}
\end{figure}
\subsection{LSE Fine Estimation} \vspace{-0.0mm}

Once the positions of the non-zero have been detected,  the next step is to estimate the value of coefficient matrix $\mathbf{U}(h) $. For the problem, we propose a novel strategy based on LSE method. This strategy is to swap the position of $h_{ij}$ and $ b_{ij}$ in the (17), so that we can get an accurate estimation by leveraging sparsity of $\textbf{b}$. Rewriting the (17) as:
\begin{equation}\label{25}   %%
\mathbf{\bar{y}} =\mathbf{\bar{S}}\mathbf{U(b)h}+\mathbf{\bar{n}}
\end{equation}
where $\mathbf{h} \in \mathbb{C}^{N_tN_r \times 1 }$. Similar with Coarse Estimation, the estimation of $\mathbf{h}$ can be found by solving the following LS problem:
\begin{equation}\label{26}   %%
\mathbf{\hat{h}} = arg \min_{\tilde{\mathbf{h}}}\{ \| \mathbf{\bar{y}}-\mathbf{\bar{S}}\mathbf{U(\hat{b}})\mathbf{\tilde{h}}\|^{2}_2\}
\end{equation}
where $ \mathbf{\hat{b}}$ is the vector obtained by SMP algorithm. Calculating the gradient of the above expression with respect to $ \tilde{\mathbf{h}}$ and setting it to zero, we get the following estimator for $ \tilde{\mathbf{h}}$ :
\begin{equation}\label{27}   %%
\left\{ \begin{array}{l}
\mathbf{\hat{h}}(k) %= \Mathbf{\Hat{Q}}^{\Dagger}_B
%\Mathbf{U}(\Mathbf{\Hat{B}(K-1)})\Mathbf{\Bar{S}}^T\Mathbf{\Bar{Y}}
%=\mathbf{\hat{Q}}^{\dagger}(k-1)(\mathbf{\bar{S}}\mathbf{U}(\mathbf{\hat{b}(k-1)}))^T\mathbf{\bar{y}} \\
%\mathbf{V}_h(k)=Vari(\hat{\mathbf{h}}(k))=\sigma^{2}_n(\mathbf{\bar{S}}\mathbf{U}(\mathbf{b}(k-1)))^T \mathbf{\bar{S}}\mathbf{U}(\mathbf{b}(k-1))
=\mathbf{\hat{Q}}^{\dagger}(k-1)(\mathbf{\bar{S}}\mathbf{U}(\mathbf{\hat{b}(k-1)}))^T\mathbf{\bar{y}} \\
\mathbf{V}_h(k)%=Vari(\hat{\mathbf{h}}(k))
=\sigma^{2}_n(\mathbf{\bar{S}}\mathbf{U}(\mathbf{\hat{b}}(k-1)))^T \mathbf{\bar{S}}\mathbf{U}(\mathbf{\hat{b}}(k-1))
\end{array} \right.\quad
\end{equation}
where $\mathbf{\hat{Q}}(k-1)=\mathbf{U}(\mathbf{\hat{b}}(k-1))\mathbf{\bar{S}}^T\mathbf{\bar{S}}\mathbf{U}(\mathbf{\hat{b}}(k-1)) $, $ \mathbf{V}_h=[v_{h_{im}}]_{N_rN_t \times 1},i\in \{1,2,...,N_r\} $ and $ m \in \{1,2,...,N_t\}$. $ \mathbf{\hat{h}(k)} $ and $\mathbf{V}_h $  denotes the estimated value and variance of $ \mathbf{h}$ at $k$th iteration. After we obtain $ \mathbf{\hat{h}}(k) $, this value will replace the $\mathbf{\hat{h}}(k-1) $ for calculating $ \mathbf{\hat{b}}(k) $ by applying SMP algorithm. \vspace{-0mm}
\subsection{ Decision and Output of LSE-SMP }

When the MSE of the LSE-SMP meets the requirement or the number of iterations reaches the limit, we output: \vspace{-0mm}
\begin{equation}\label{28} \vspace{-0mm}
\hat{b}_{ij}= \frac{1}{1+\frac{\prod_{t=1}^{T}(1-p^{s}_{it \rightarrow tj}(k))\cdot p_0(b_{ij}=0)}{\prod_{t=1}^{T}p^{s}_{it \rightarrow tj}(k)\cdot p_0(b_{ij}=1)}}
\end{equation}
and
\begin{equation}\label{29}
\left\{ \begin{array}{l}
\hat{\mathbf{h}}=\mathbf{\hat{Q}}^{\dagger}(\mathbf{\bar{S}}\mathbf{U}(\mathbf{\hat{b}}))^T\mathbf{\bar{y}},\\
\mathbf{\hat{h}}_{v}^*=\mathbf{U}(\mathbf{\hat{h}})\mathbf{\hat{b}}.
\end{array} \right.\quad
\end{equation} \vspace{-0mm}
where $\mathbf{\hat{h}}_{v}^*$ is the finally outputted channel estimation vector. It should be pointed out that the decision is made based on full information coming from all the sum nodes.
\section{ANALYSIS OF CRAMER RAO BOUND }
%The Cramer-Rao Lower Bound (CRLB) serves as an important tool in the performance evaluation of estimators which arise frequently in the fields of communications
%and signal processing. In terms of proposed sparse estimation algorithm,
In this section, we give the analysis of Cramer-Rao bound and show that our proposed LSE-SMP algorithm is unbiased. We begin with conventional LSE that is applied for a deterministic and non-sparse channel vector $\mathbf{h}_v$. From \cite{CRLB_book} and recalling the signal model in (13), the CRLB is yielded as:\vspace{-0.0cm}
\begin{equation}\label{21}
\mathbf{CRLB_{LSE}}\geq \sigma^{2}_n(\mathbf{\bar{S}}^T \mathbf{\bar{S}})^{-1}
\end{equation} \vspace{-0.0cm}
Note that LSE is the Minimum Variance Unbiased Estimator (MVUE), if $\mathbf{h}_v$ is deterministic and non-sparse. %It can be easily shown that $E(\mathbf{\hat{h}_v})=E\{ (\mathbf{\bar{S}}^T\mathbf{\bar{S}})^{-1}\mathbf{\bar{S}}^T \mathbf{\bar{y}}\}=\mathbf{h}_v $. Detailed proof can be found in \cite{CRLB_book}.

Compared with the non-sparse case, the sparse situation is slightly more complicated. As we are interested in the lower bound for the estimation accuracy, we assume perfect knowledge of the non-zero positions, i.e. $ \mathbf{U}(\mathbf{\hat{b}})=\mathbf{b} $. Then, the next step is to verify the estimator is unbiased. Recalling the signal model in (25) and from the definition of unbiased estimator, we get:
%\begin{equation}  %%
\begin{align}
\!\!E(\mathbf{\hat{h}}_v) & \!=\!\!E(((\mathbf{\bar{S}}\mathbf{U}(\mathbf{b}))^T\mathbf{\bar{S}}\mathbf{U}(\mathbf{b}))^\dag(\mathbf{\bar{S}}\mathbf{U}(\mathbf{b}))^T \mathbf{\bar{y}}) \nonumber \\
& \!\!=\!\!E(((\mathbf{\bar{S}}\mathbf{U}(\mathbf{b}))^T\mathbf{\bar{S}}\mathbf{U}(\mathbf{b}))^\dag(\mathbf{\bar{S}}\mathbf{U}(\mathbf{b}))^T (\mathbf{\bar{S}}\mathbf{U}(\mathbf{b}) \mathbf{h}+\mathbf{\bar{n}})) \nonumber \\
&\!\!=\!\!\mathbf{U}(\mathbf{b})\mathbf{h}=\mathbf{h}_v
\end{align}  \vspace{-0.3cm}
%\end{equation}

So, it is a unbiased estimator. The next step is to compute its CRLB. As previously mentioned, the channel  $ \mathbf{h}_v $ is a deterministic vector, we can get:
\begin{equation}\label{21}   %%
\mathbf{\bar{y}} \sim p(\mathbf{\bar{y}};\mathbf{h}_v)=\mathcal{N}(\mathbf{\bar{S}}\mathbf{h}_v,\sigma^{2}_n\mathbf{I})
\end{equation}

Then, we can compute the $ \frac{ \partial ln \,p(\mathbf{\bar{y}};\mathbf{h}_v)}{\partial \mathbf{h}_v } $,
\begin{equation}\label{21}   %%
%\left\{
%\begin{array}{l}
\!\!\!\!\frac{ \partial ln \,p(\mathbf{\bar{y}};\mathbf{h}_v)}{\partial \mathbf{h}_v }
%= \frac{\partial (-ln\,(\sqrt{2\pi}\sigma_n)-\frac{1}{2\sigma_n^{2}}(\mathbf{\bar{y}}-\mathbf{\bar{S}}\mathbf{h}_v)^T(\mathbf{\bar{y}}-\mathbf{\bar{S}}\mathbf{h}_v))}{\partial \mathbf{h}_v } \\
   \!\! =\!\!\frac{1}{\sigma_n^{2}}[(\mathbf{\bar{S}}\mathbf{U}(\mathbf{b}))^T\mathbf{\bar{y}}-(\mathbf{\bar{S}}\mathbf{U}(\mathbf{b}))^T \mathbf{\bar{S}}\mathbf{U}(\mathbf{b})\mathbf{h}_v]
%\end{array}
%\right.
\end{equation}

Then, we obtain the following expression for the Fisher Information Matrix (FIM):
\begin{equation}\label{21}   %%
I(\mathbf{h}_v)=-E(\frac{ \partial^2 ln\, p(\mathbf{\bar{y}};\mathbf{h}_v)}{\partial \mathbf{h}^{2}_v}) \\
           = \frac{1}{\sigma_n^{2}}{{(\mathbf{\mathbf{\bar{S}}}\mathbf{U}(\mathbf{b}))^T\mathbf{\bar{S}}}\mathbf{U}(\mathbf{b})}
\end{equation}

We note that $ \mathbf{\bar{S}} \mathbf{U}(\mathbf{b}) $ has the rank no larger than $L$  due to the multiplication of $ \mathbf{\bar{S}} $ by $ \mathbf{U}(\mathbf{b})$. The matrices $ \mathbf{\bar{S}} \mathbf{U}(\mathbf{b}) $ and $(\mathbf{\bar{S}} \mathbf{U}(\mathbf{b}))^T $ have some all zero columns (and rows), so it is singular. For this type singular matrix, it need to meet the following constraint \cite{CRLB_kay}, otherwise our proposed estimator (29) has infinite variance that renders the CRLB useless. Before we analyze this constraint, we firstly compute the following key identity:
\begin{equation}\label{21}   %%
%\left\{
%\begin{array}{l}
\begin{split}
\mathbf{G}
%&=\frac{\partial(E\{\hat{\mathbf{h}}_v-\mathbf{h}_v\})}{\partial \mathbf{h}_v^T}-\frac{\partial \mathbf{h}_v }{\partial \mathbf{h}_v^T} \\
&=((\mathbf{\bar{S}}\mathbf{U}(\mathbf{b}))^T\mathbf{\bar{S}}\mathbf{U}(\mathbf{b}))^\dag(\mathbf{\bar{S}}\mathbf{U}(\mathbf{b}))^T\mathbf{\bar{S}}\mathbf{U}(\mathbf{b}) \\
&= diag(\mathbf{b})\neq \mathbf{I}_{N_rN_t}
%\end{array}
%\right.
\end{split}
\end{equation}

This constraint is given by:
\begin{equation}\label{21}   %%
\mathbf{G}=\mathbf{G}I(\mathbf{h}_v)I(\mathbf{h}_v)^\dag
\end{equation}
plugging  the (34) and (35) into (36), we obtain $ \mathbf{G}=\mathbf{G}I(\mathbf{h}_v)I(\mathbf{h}_v)^\dag = \mathbf{U(b)U(b)}$ that holds. This means that the variance of our proposed estimator is finite.
Since the FIM $ I(\mathbf{h}_v) $ in (34) is  singular, the expression for the CRLB can be computed following \cite{CRLB_kay} which yields:
\begin{equation}\label{21}   %%
%\left\{
%\begin{array}{l}
\begin{split}
\mathbf{CRLB_{LSE-SMP}} & \geq \mathbf{G}I(\mathbf{h}_v)^\dag \mathbf{G}^T \\ %=\sigma^{2}_n\mathbf{G}(\mathbf{\bar{S}}\mathbf{U}(\mathbf{b}))^T\mathbf{\bar{S}}\mathbf{U}(\mathbf{b})\mathbf{G}^T \\
& =\sigma^{2}_n(\mathbf{\bar{S}}\mathbf{U}(\mathbf{b}))^T\mathbf{\bar{S}}\mathbf{U}(\mathbf{b})
%\end{array}
%\right.
\end{split}
\end{equation}

%Assume that the pdf $ ln \, p(\mathbf{\bar{y}};\mathbf{h}_v) $  satisfies the  regularity condition:
%\begin{equation}\label{21}   %%
%E[\frac{ \partial ln \,p(\mathbf{\bar{y}};\mathbf{h}_v)}{\partial \mathbf{h}_v } ]     =  0, \\
%\end{equation}
%
%An unbiased estimator that attains the CRLB can be found iff:
%\begin{equation}\label{21}   %%
%\frac{ \partial ln \,p(\mathbf{\bar{y}};\mathbf{h}_v)}{\partial \mathbf{h}_v }    =  I(\mathbf{h}_v)(\mathbf{h}_{\mathbf{CRLB}}-\mathbf{h}_v)  \\
%\end{equation}
%
%The combination of (33) and (38) yields the following key results:
%\begin{equation}\label{21}   %%
%\left\{
%\begin{array}{l}
%\hat{H}_v=((\mathbf{\bar{S}}\mathbf{U}(\mathbf{b}))^T\mathbf{\bar{S}}\mathbf{U}(\mathbf{b}))^\dag(\mathbf{\bar{S}}\mathbf{U}(\mathbf{b}))^T \mathbf{\bar{y}}    \\
%\mathbf{CRLB_{LSE-SMP}}=\mathbf{I}^{\dag}(\mathbf{h}_v)=\sigma^{2}_n(\mathbf{\bar{S}}\mathbf{U}(\mathbf{b}))^T\mathbf{\bar{S}}\mathbf{U}(\mathbf{b})
%\end{array}
%\right.
%\end{equation}
%Compared with (29) and (39), we find that they are consistent results. This shows that our proposed LSE-SMP estimators can be the MVUE under the condition that has knowledge of the non-zero positions in the channel vector.
%the channel matrix $\mathbf{ H_v}$ that is a deterministic and we have the .
\section{NUMERICAL RESULTS}
In this section, we  investigate the performance of the our proposed algorithm using Monte-Carlo simulations, comparing with the LASSO sparse estimator and CRLB. We demonstrate that ours proposed LSE-SMP algorithm provides significant performance gains over existing techniques. In particular, we show that our proposed algorithm specialize in the mmWave channel estimation. Furthermore, we conduct numerical studies to investigate the impact of channel sparse ratios and iterations.

\subsection{Setup}
For our numerical study, we considered the channel estimation problem in a $ 32 \times 64 $ mmWave MIMO system. The value and positions of non-zero elements in original channel vector $ \mathbf{h}_v $ were both generated by random way with a distribution following $\mathcal{N}(0,\sqrt{10}) $. Throughout, we considered SNR $\triangleq E\{\|\mathbf{\bar{S}} \|^{2}_F / \|\mathbf{\bar{n}} \|^{2}_F\}$ in the interval [-10, 40dB]; sparsity ratio was calculated by $\eta= L/(N_r \times N_t)$; and the performance metric was the  Normalized Mean Square Error (NMSE), given by $ E\{\frac{ \| \mathbf{\hat{h}}_{v}^*-\mathbf{h}_{v}\|^{2}_F}{\| \mathbf{h}_{v}\| ^{2}_F}\}$. \vspace{-1mm}
\begin{figure}
  \begin{center}
  \includegraphics[width=76mm]{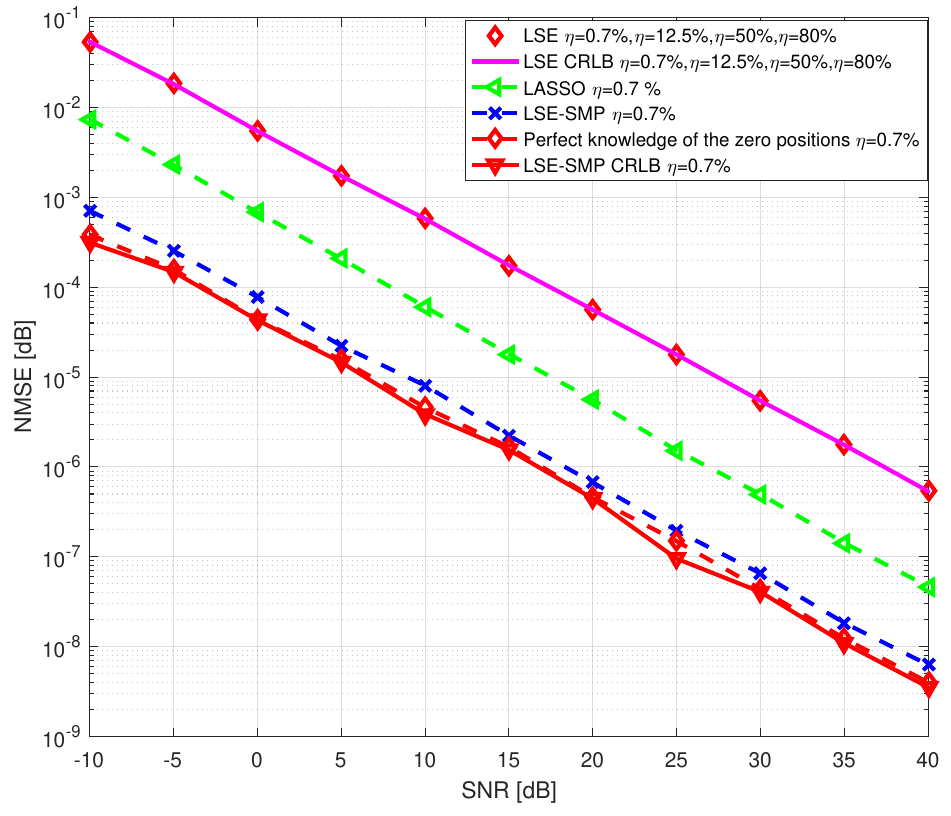}\\ \vspace{-4mm} %   %\captionsetup{justification=centering}
  \caption{Average NMSE performance comparison of proposed LSE-SMP, conventional LSE, and LASSO channel estimates versus SNR for the sparsity ratio at $\eta =0.7\% $ with the 5 iterations. It also shows CRLB of LSE and proposed LSE-SMP.}
  \label{fig:env} \vspace{-7.5mm}
  \end{center}
\end{figure}
\subsection{Performance Comparison}
 Fig. 4 shows the average channel-estimation NMSE of our proposed LSE-SMP algorithm and LASSO under the sparse ratio at $0.7\% $. Additionally, we also compute the CRLB for the classical LSE and our proposed LSE-SMP estimator. Our proposed LSE-SMP estimator obtains the better NMSE performance than that of LASSO and LSE. As expected, the CRLB for our proposed LSE-SMP is the lowest, and this result is also consistent with that of classical LSE estimator with perfect knowledge of the non-zero positions. It is also seen from the Fig.4 that the gap between LSE-SMP and LSE-SMP CRLB is much smaller, about $1.5dB$. This is partly due to the errors in the detection of the non-zero positions in sparse message passing phase and partly to the fact that all our detection strategies rely on a coarse (and noisy) initial estimate of the channel.\vspace{-0.2cm}
\subsection{Effect of Sparsity Ratio}

For further investigating the effect of sparsity ratio to our proposed algorithm, we fixed the iteration times $=5$, and changed the sparsity ratio $\eta$ from $0.7\%$ to $ 80\%$. Note that different sparsity ratios $ \eta $ will leads to different LSE-SMP CRLBs.  Simulations show that the NMSE performance of the LSE-SMP is very close to its CRLB in the each of sparsity ratios. In the Fig. 5, we just give the LSE-SMP CRLB at the sparse ratio $ \eta=0.7\%$. We can see that the NMSE performance of LSE is consistent LSE CRLB, and it is also the upper bound of LSE-SMP. These verify the analysis of LSE-SMP in the IV section. It means that the NMSE performance of LSE-SMP will be better with decreasing of sparse ratios as shown in the Fig. 5, essentially because LSE-SMP is able to exploit the sparsity of the channel. To be specific, The NMSE performance of LSE keep unchange under different sparsity ratios. On the other hand, the NMSE performance of LSE-SMP will increase with the decreasing of sparsity ratios. This means that LSE-SMP scheme will perform better especially when channel is very sparse.

\begin{figure}
  \begin{center}
  \includegraphics[width=78mm]{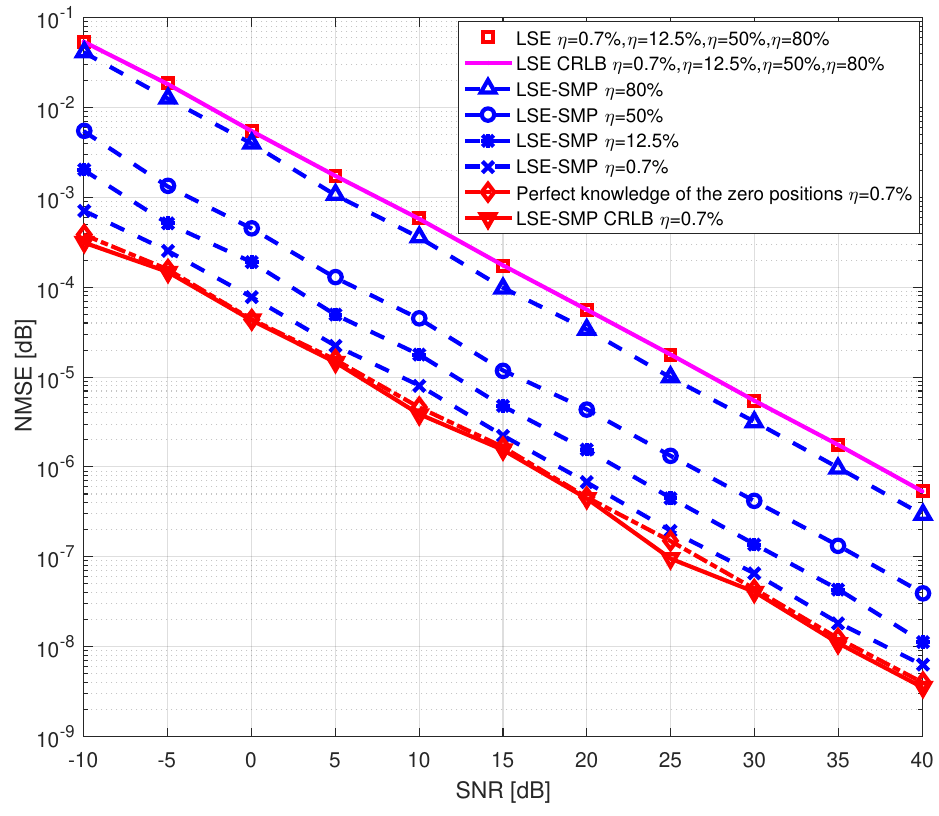} \vspace{-4mm}\\  %   %\captionsetup{justification=centering}
  \caption{Average NMSE performance comparison of LSE-SMP, LSE and their CRLB channel estimates versus SNR for different sparsity ratios $\eta \in \{ 0.7 \%,12.5 \%,50 \%,80 \%\} $ with the iteration $=5$. It can be seen that LSE CRLB is the upper bound and LSE-SMP CRLB is the low bound of our proposed LSE-SMP.}
  \label{fig:env} \vspace{-7.5mm}
  \end{center}
\end{figure}

\subsection{Effect of Iterations }

Fig. 6 shows the average channel-estimation NMSE performance for LSE-SMP algorithm under several turbo iterations with sparsity ratio $\eta=3.1 \%$. The result shows that after the second turbo iteration, the NMSE performance of LSE-SMP performs a significant improvement. Additionally, we also find that the gap of the NMSE performance between the adjacent iterations for LSE-SMP algorithm will be decreasing with increasing of iterations. After fourth turbo iteration, the NMSE performance have no significant improvement and it is very close to our analyzed LSE-SMP CRLB. This demonstrates that the convergence speed of the LSE-SMP algorithm is fast.

\section{CONCLUSION}

In this paper, a novel channel estimation algorithm for mmWave MIMO systems was proposed, which can take full advantage of the inherent sparseness of mmWave channel. This algorithm leverages the virtues of the SMP and LSE algorithm. The CRLB for the proposed LSE-SMP algorithm was analyzed. Simulation experiments verified that our proposed algorithm had much better performance than the existing sparse estimators, especially when channel is very sparse. Furthermore, it was also shown that the proposed algorithm needs only several turbo iterations to achieve CRLB. In spite of this, it still has the high complexity due to the computation of inverse matrixes in coarse and fine estimation phases. In the following, we plan to develop an approach to replace the LSE coarse and fine estimation without loss of global performance of the algorithm. Another limitation in our paper is the assumption that virtual channel vector $\mathbf{h}_v$ has exactly $L$ non-zero entries. In the future work, we will also consider to relax the assumption.
\begin{figure}
  \begin{center}
  \includegraphics[width=78mm]{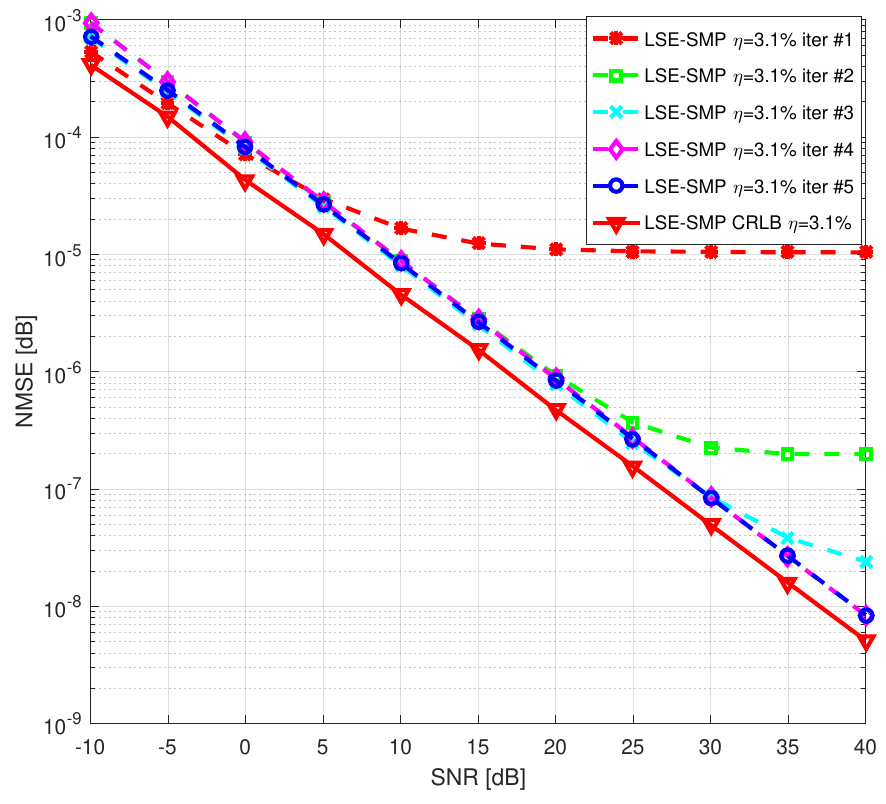} \vspace{-4mm}\\  % %\captionsetup{justification=centering}
  \caption{Average NMSE performance comparison of LSE-SMP and its CRLB channel estimates versus SNR for different turbo iterations with the sparsity ratio $\eta=3.1 \%$. After the fourth iteration, the NMSE performance of LSE-SMP is converging to its CRLB. }
  \label{fig:env} \vspace{-7.5mm}
  \end{center}
\end{figure}

\end{document}